\documentclass[conference]{IEEEtran}
\IEEEoverridecommandlockouts
\usepackage{cite}
\usepackage{amsmath,amssymb,amsfonts}
\usepackage{graphicx}
\usepackage{textcomp}
\usepackage{xcolor}
\usepackage{algpseudocode}
\usepackage{subcaption}
\usepackage[ruled,vlined]{algorithm2e}
\def\BibTeX{{\rm B\kern-.05em{\sc i\kern-.025em b}\kern-.08em
    T\kern-.1667em\lower.7ex\hbox{E}\kern-.125emX}}

\begin{document}

\newtheorem{proposition}{Proposition}

\title{{Spatial Signal Strength Prediction using 3D Maps and Deep Learning}\\ 
\thanks{This work was supported in part by the CONIX Research Center, one of six centers in JUMP, a Semiconductor Research Corporation (SRC) program sponsored by DARPA.}
}

\author{
    \IEEEauthorblockN{Enes~Krijestorac, Samer Hanna, Danijela Cabric}
    \IEEEauthorblockA{\textit{Electrical and Computer Engineering Department,} \textit{University of California, Los Angeles}, USA
    \\enesk@ucla.com, samerhanna@ucla.edu, danijela@ee.ucla.edu} 

}

\maketitle

\begin{abstract}
Machine learning (ML) and artificial neural networks (ANNs) have been successfully applied to simulating complex physics by learning physics models thanks to large data. 
Inspired by the successes of ANNs in physics modeling, we use deep neural networks (DNNs) to predict the radio signal strength field in an urban environment. Our algorithm relies on samples of signal strength collected across the prediction space and a 3D map of the environment, which enables it to predict the scattering of radio waves through the environment. While already extensive body of research exists in spatial signal strength prediction, our approach differs from most existing approaches in that it does not require the knowledge of the transmitter location, it does not require side channel information such as attenuation and shadowing parameters, and it is the first work, to the best of our knowledge, to use 3D maps to accomplish the task of signal strength prediction. This algorithm is developed for the purpose of placement optimization of a UAV or mobile robot to maximize the signal strength to or from a stationary transceiver but it also holds relevance to dynamic spectrum access networks, cellular coverage design, power control algorithms, etc. 
\end{abstract}


\section{Introduction}
Mobile devices such as unmanned aerial vehicles (UAVs) or ground robots can optimize their location in order to maximize the signal strength to a basestation or user equipment on the ground. Various placement optimizations algorithms for signal strength have been proposed to solve for optimal placement of such UAVs or mobile robots. We can roughly divide the said placement optimization algorithms into model-free and model-based solutions. In model-free solutions the UAV does not aim to predict the signal strength across the entire optimization space but rather finds a path that maximizes the expected increase in signal strength \cite{ladosz2016optimal, krijestorac}. Model-based solutions rely on being able to predict the signal strength across the optimization space and use that to find the optimal path to maximize the signal strength \cite{jin2012joint, esrafilian2018uav}. 
At the heart of model-based solutions is some type of spatial signal strength prediction algorithm, which relies on some prior information, such as transmitter location, previously collected signal strength measurements, estimated parameters of channel attenuation and shadowing. The use of spatial signal strength prediction algorithms extends beyond placement of mobile radio devices. They are widely used to enable dynamic spectrum access in cognitive radio networks, improve cellular coverage, and facilitate power control. 

Given the growing importance of spatial signal strength prediction algorithms, we address the limitations of algorithms previously developed for this task. Namely, we propose a novel deep-learning-based spatial signal strength prediction algorithm that differs in three key aspects compared to most existing approaches: 1) it does not rely on the knowledge of the location of source transmitter; 2) it does not rely on the knowledge of the parameters of the shadowing and path loss models and lastly, and 3) it incorporates the knowledge of the 3D map of the environment; 3D maps allow the algorithm to predict scattering and blockage due to the buildings. 
Like the existing prediction algorithms in literature, our proposed algorithm relies on a small number of measurements collected across a target space to predict the signal strength across the remaining locations. 
DNNs are known to be universal function approximators, which is why we use their capabilities to learn the complicated relationship between 3D objects, signal strength measurements and the entire signal strength field. 
Another important feature of our algorithm is that rather than simply predicting estimates of signal strength across the target space, our algorithm models the signal strength as a Gaussian random variable and predicts its mean and variance. Knowing the signal strength prediction variance is particularly important for path planning algorithms for UAVs and mobile robots.

We briefly review and categorize previously developed signal strength prediction algorithms. The most common approaches for prediction are adopted from the field of spatial interpolation \cite{angjelicinoski2011comparative, chowdappa2018distributed, hernandez2012field, braham2016spatial}. Among the interpolation methods, the most common ones are inverse distance weighting (IDW), gradient plus inverse distance squared (GIDS) and Kriging interpolation, which relies on spatial statistics. Algorithms based on Kriging interpolation rely on the location of the source, while IDW or GIDS based algorithms normally do not. Kriging approach is the most commonly used spatial interpolation method since it can be tailored specifically to the statistical models of wireless channel shadowing. 
Furthermore, a variety of approaches were developed based on Bayesian inference, which normally rely on statistical channel models and source location as well \cite{romero2020aerial, sayrac2013bayesian, muppirisetty2015spatial, huang2014cooperative}. Other stand-alone approaches were proposed based on low rank and sparse matrix reconstruction \cite{lee2017channel} and MMSE estimation \cite{malmirchegini2012spatial}. Furthermore, deep learning (DL) has also been considered in literature recently as of the time of this writing \cite{han2020power, teganya2020deep}. In \cite{han2020power}, generative adversarial neural networks are used and the authors in \cite{teganya2020deep} use deep completion auto-encoders. Like our approach, the cited DL approaches do not rely on user location or statistical channel models and its parameters. However, the DL algorithm we are proposing differs in that it outputs a probability distribution over inferred signal strength and in that it can utilize 3D map information.      

The remainder of this paper is organized as follows. In Sec. \ref{sec:sys-model} we define the system model. In Sec. \ref{sec:approach}, we explain our proposed approach and its basic building blocks. In Sec. \ref{sec:dataset} we explain how we obtained the dataset to train our algorithm and the details of the training. In Sec. \ref{sec:results} we showcase our results and provide our explanations for them. Finally, in Sec. \ref{sec:conclusions}, we summarize this paper and the main outcomes while also addressing the future steps. 


\section{System Model}
\label{sec:sys-model}
\begin{figure}[t]
	\centering
	\includegraphics[width=0.28\textwidth]{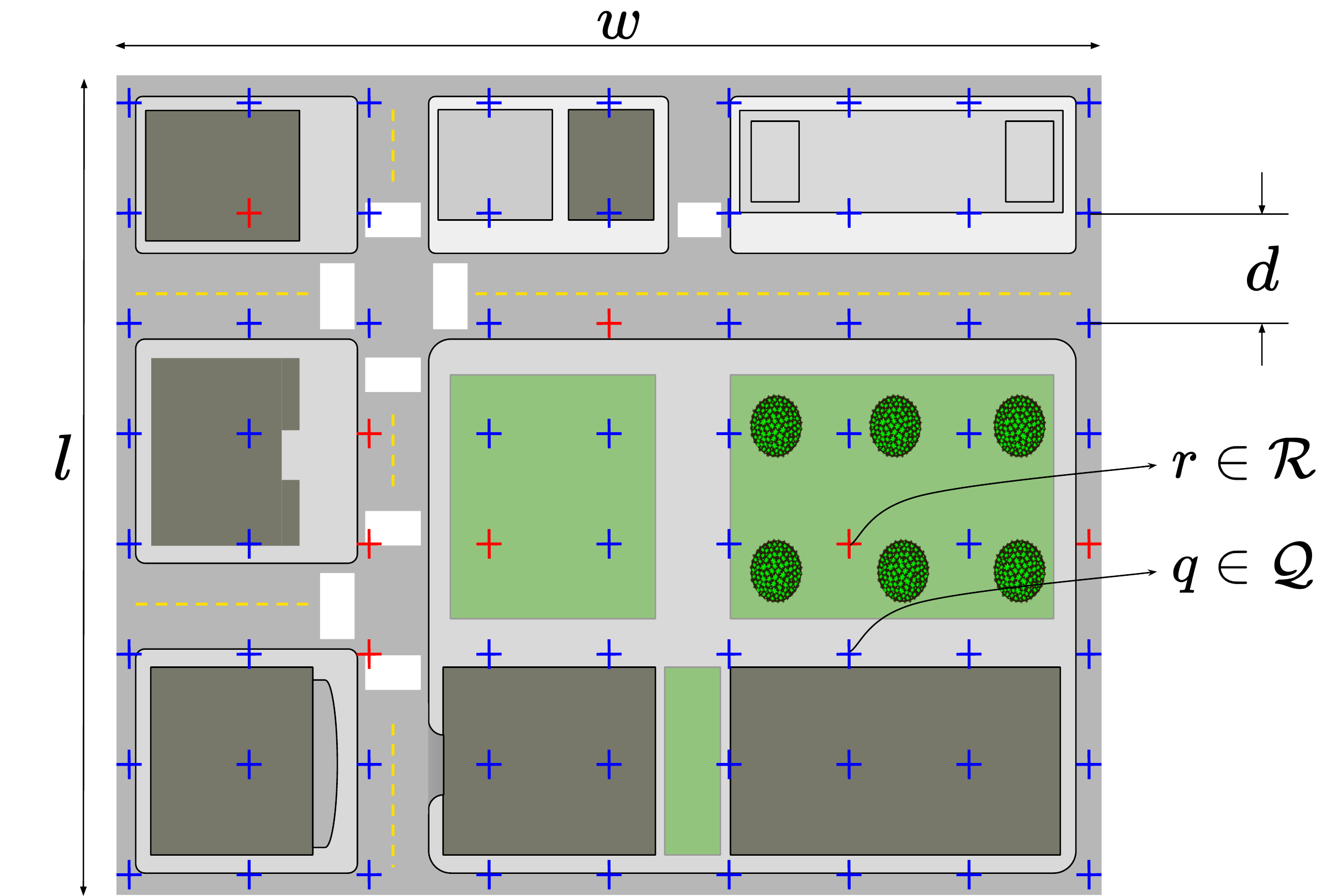}
	\caption{Overview of the parameters and variables. The set of coordinates $\mathcal{R}$ refers to the locations where signal strength measurements have been collected. The set of coordinates $\mathcal{Q}$ refers to the entire set of discrete locations across the area of interest.}
	\label{fig:sys}
	\vspace{-15px}
\end{figure}
In this section we define the scenarios in which our algorithm is designed to operate, introduce operating variables and explain our assumptions.  

We consider a rectangular urban area of interest (AoI) of width $w$ and length $l$, for which we have a database of major buildings and objects that can be used to construct a 3D map of the environment, such as the one shown in Fig. \ref{fig:sys}. The coordinates of a point in the AoI are denoted by $(x,y,h)$, where $x$ and $y$ correspond to the horizontal coordinates and $h$ is the altitude. The communication occurs between an outdoor ground user (GU) at an altitude $h_{\text{GU}}$ and a UAV or a mobile robot at an altitude $h_{\text{UAV}}$. The GU is not necessarily located within the AoI. The GU is also assumed to be permanently stationary or stationary for the period of time for which we are trying to predict the signal strength over the AoI. 

Let $\gamma(x,y,h)$ be the logarithm of the channel gain between the GU and a transceiver at location $(x,y,h)$ and $P_{T,dBm}$ be the transmit power in dBm, then the received signal power at $(x,y,h)$ is $P_{T,dBm}+\gamma(x,y,h)$. We assume that the gain is reciprocal, i.e. the gain of transmission from the GU to the UAV is equal to gain of transmission from the UAV to the GU.  

In a line-of-sight (LOS) scenario, where the signal reaches the receiver along the straight line path, the gain can be accurately predicted using path-loss models and by knowing the antenna properties of the receiver and the transmitter. In a non-line-of-sight (NLOS) environment, where the received signal is an addition of the straight line paths and the reflected and diffracted paths, prediction of the gain becomes challenging. The most accurate tools for prediction of the gain in such environments are ray-tracing software which use 3D maps of the transmission environment to simulate all possible paths between the transmitter and the receiver. 
However, ray tracing software is computationally expensive and 3D maps of the transmission environments are generally not available, which is why researchers and engineers normally resort to statistical models of the gain. In such models, the gain $\gamma(x,y,h)$ is split into three components, $\gamma(x,y,h) = \psi_{PL}(x,y,h)+\psi_{SH}(x,y,h)+\psi_{F}(x,y,h)$, where $\psi_{PL}(x,y,h)$ is the path loss due free space attenuation, $\psi_{SH}(x,y,h)$ is the loss due to shadowing and $\psi_{F}(x,y,h)$ is the loss due to random fading. While $\psi_{PL}(x,y,h)$ can be accurately predicted knowing the antenna properties of the receiver and the transmitter, $\psi_{SH}(x,y,h)$ and $\psi_{F}(x,y,h)$ are modelled as random variables with some appropriately fitted probability distributions. $\psi_{SH}(x,y,h)$ is often modelled as correlated over time and space, while $\psi_{F}(x,y,h)$ is modelled as an i.i.d. random variable. For example, $\psi_{SH}(x,y,h)$ is often modeled as normal random variable with exponentially decaying spatial correlation according to the Gudmundson model \cite{gudmundson1991correlation}.  


Our main objective is the prediction of the signal strength between the GU and the UAV based on previously collected measurements of signal strength across the AoI and 3D map of the AoI. The measurements may have been collected by the UAV itself as it explores the space and/or by a number of sensors that collect the measurements across the space. Our algorithm does not rely on any statistical models of the channel. Rather, we use DL models to capture the relationship between measurements of signal strength, 3D map of the transmission environment and the expected signal strength at a location in the AoI. 

We focus on predicting the signal strength at the altitude of the UAV $h_U$, and we assume that the measurements of signal strength that are used to inform the prediction are collected at the same altitude. Our approach is demonstrated for 2D, however extension to 3D is viable with a more complex DL architecure and with more data. The space is discretized into a grid with grid points equally spaced by separation $d$ as shown in Fig. 1. We denote the coordinates of a grid point as $q$ and the set of all grid points in the AoI is $\mathcal{Q}$ ($|\mathcal{Q}|=\frac{wl}{d^2}$). Furthermore, the coordinates of grid points at which measurements have been collected are denoted as $r$ and the set of all such coordinates is denoted as $\mathcal{R}$. The measured signal strength values in dBm at coordinates $r \in \mathcal{R}$ are denoted as $\mathbf{x} \in \mathbb{R^{|\mathcal{R}|}}$ and the 3D map of the environment is denoted as $\mathcal{M}$. The signal strength values in dBm at all coordinates are denoted as $\mathbf{y} \in \mathbb{R^{|\mathcal{Q}|}}$. The signal strength at coordinates that are obstructed by the buildings is naturally zero. We use an indicator binary variable $\mathbf{z} \in \mathbb{Z}_2^{|\mathcal{Q}|}$, where $[\mathbf{z}]_j=0$ if the location $q_j$ is obstructed by a building, i.e. it is indoor, and $[\mathbf{z}]_j=1$ is outdoor. 

\section{Proposed approach}
\label{sec:approach}
In this section we explain the reasoning behind our approach and the details of its building blocks. 

\subsection{Intuition}


Inspired by the successes of neural networks in physics modeling, we apply neural networks to modelling of the propagation of radio signals. %
While radio propagation in an urban environment can be modeled using ray tracing software, neural network can be trained to accomplish the same task while requiring far less computing power.   
Moreover, unlike a ray tracing software, our algorithm does need to know the location of the transmitter. Instead, our algorithm relies on previously collected signal strength measurements and 3D maps to accomplish signal strength prediction without knowing user locations.

\subsection{Algorithm and loss function}

We now describe our deep learning algorithm and the loss function used to train it.   

Let $\mathbf{\hat{y}}$ denote the predicted signal strength value. Instead of simply predicting the expected value of $\mathbf{\hat{y}}$, we predict a probability distribution of $\mathbf{\hat{y}}$, $f(\mathbf{\hat{y}}|\mathbf{x}, \mathcal{M})$. Predicting a probability distribution of $\mathbf{\hat{y}}$ rather than just the expected value of $\mathbf{\hat{y}}$ is important for path planning algorithms that would guide a UAV or a mobile robot to a location of maximum or sufficient signal strength. 
Based on the Gudmundson model of shadowing, we chose Gaussian distribution with mean $\mathbf{\mu}$ and covariance $\mathbf{\Sigma}$ as a fit for $f(\mathbf{\hat{y}}|\mathbf{x}, \mathcal{M})$. 
Our algorithm learns the mappings from $(\mathbf{x}, \mathcal{M})$ to $\mathbf{\mu}$ and $\mathbf{\Sigma}$, which we denote as $\mathbf{\mu}(\mathbf{x}, \mathcal{M})$ and $\mathbf{\Sigma}(\mathbf{x}, \mathcal{M})$. 
The function $\mathbf{\mu}(\mathbf{x}, \mathcal{M})$ is parametrized by a DNN $\mathbf{\mu}_\theta(\mathbf{x}, \mathcal{M})$ with parameters $\theta$ and the function $\mathbf{\Sigma}(\mathbf{x}, \mathcal{M})$ is parametrized by a DNN $\mathbf{\sigma}_\omega^2(\mathbf{x}, \mathcal{M})$ with parameters $\omega$. In order to reduce the complexity of the neural network, we assume that $\mathbf{\Sigma}(\mathbf{x}, \mathcal{M})$ is a diagonal matrix and so $\mathbf{\sigma}_\omega^2(\mathbf{x}, \mathcal{M}) \in \mathbb{R}^{|\mathcal{Q}|}$. We denote the fitted Gaussian distribution as $f_{\theta, \omega}(\mathbf{\hat{y}}|\mathbf{x}, \mathcal{M})$.

The loss function is based on Kullback–Leibler (KL) divergence between $f(\mathbf{\hat{y}}|\mathbf{x}, \mathcal{M})$ and $f_{\theta, \omega}(\mathbf{\hat{y}}|\mathbf{x}, \mathcal{M})$,
\begin{multline}
    \mathcal{L}(\theta, \omega) = 
    \frac{1}{2N}\sum_{i=1}^N\Delta_i^T 
    \text{diag}\left(\mathbf{z} \right) 
    \text{diag}\left(\mathbf{\sigma}^2_\omega(\mathbf{x}_i, \mathcal{M}_i)  \right)^{-1}
    \Delta_i   + \\
     \frac{1}{2N} \left( \mathbf{1}^T \left( \log(\text{diag}(\mathbf{z})  \sigma^2_\omega(\mathbf{x}_i)) \right) \right)
\end{multline}
where $N$ is the number of training samples. Each training sample consists of $\mathbf{x}$, $\mathcal{M}$ and $\mathbf{y}$. For the ease of exposition, we introduced a substitute variable $\Delta = \mathbf{\mu}_\theta(\mathbf{x}, \mathcal{M})-\mathbf{y}$. We modified the KL divergence loss function to only include error for the outdoor coordinates by including $\mathbf{z}$ in the expression. This variable is required during training but not for prediction during run time. The loss function is minimized to train the algorithm: $(\theta, \omega) = \arg\min_{\theta, \omega}\mathcal{L}$. The minimization is performed using stochastic gradient descent algorithm, as is common for training deep-learning models. 

\subsection{DNN design}
In this subsection we describe the DNN blocks $\mathbf{\mu}_\theta(\mathbf{x}, \mathcal{M})$ and $\mathbf{\sigma}^2_\omega(\mathbf{x}, \mathcal{M})$. The design of the two DNN blocks is identical, however the parameters $\theta$ and $\omega$ will be different since $\mathbf{\mu}_\theta(\mathbf{x}, \mathcal{M})$ learns to output the mean and $\mathbf{\sigma}^2_\omega(\mathbf{x}, \mathcal{M})$ the variance of $f_{\theta, \omega}(\mathbf{\hat{y}}|\mathbf{x}, \mathcal{M})$. The two DNNs are composed of three different blocks:
\begin{equation*}
    \mathbf{\mu}_\theta(\mathbf{x}, \mathcal{M})=\psi_\theta(n(t(\mathbf{x}, \mathcal{M}))) 
\end{equation*}
\begin{equation*}
    \mathbf{\sigma}^2_\omega(\mathbf{x}, \mathcal{M})=\psi_\omega(n(t(\mathbf{x}, \mathcal{M}))) 
\end{equation*}
where $t{}(\cdot)$ is a transformation block which we describe below, $n{}(\cdot)$ is a normalization block and $\psi$ is a convolutional neural network (CNN). 

The inputs $\mathbf{x}$, $\mathcal{M}$ are transformed into a 3D tensor $\tilde{\mathbf{X}}$ through the transformation $t{}(\cdot)$. Let $\mathbf{X}\in \mathbb{R}^{\frac{l}{d} \times \frac{w}{d}}$, where $\left [\mathbf{X} \right]_{i,j}$ is equal to the signal strength at $(id, jd, h_U)$, if $(id, jd, h_U) \in \mathcal{R}$, and is equal to $c_L$, otherwise. $c_L$ is a padding value that is set to a very low value outside of the reasonable range of signal strength values. Similarly, let $\mathbf{M}\in \mathbb{R}^{\frac{l}{d} \times \frac{w}{d}}$, where $\left [\mathbf{M} \right]_{i,j}$ is equal to the building height at $(id, jd)$. The matrices $\mathbf{M}$ and $\mathbf{X}$ are stacked into a 3D tensor of size $\frac{l}{d} \times \frac{w}{d} \times 2$, which we denote as $\tilde{\mathbf{X}}$. 

In $n{}(\cdot)$, the tensor $\tilde{\mathbf{X}}$ is normalized such that all entries scale to -1 to 1.

\begin{figure}[t]
	\centering
	\includegraphics[width=0.45\textwidth]{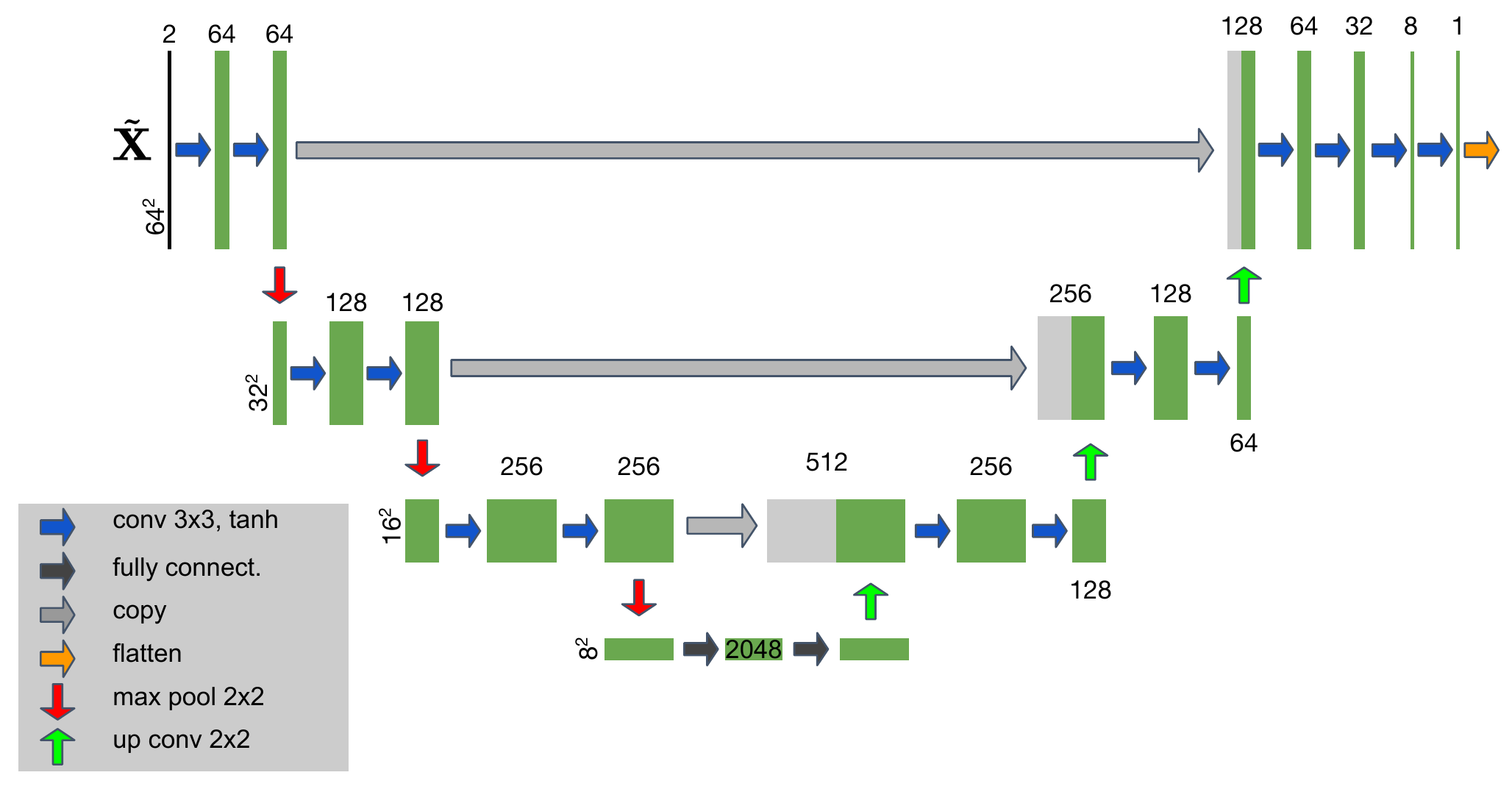}
	\caption{Overview of the U-Net CNN architecture used.}
	\label{fig:u-net}
	\vspace{-10px}
\end{figure}

We use a CNN in the DNN block since CNNs consist of spatial filters that enforce a local connectivity pattern between neurons of adjacent layers. This architecture ensures that the learned filters produce the strongest response to a spatially local input pattern. This feature is suitable for our problem since signal strength at a location is more correlated to adjacent signal strength measurements and 3D topology then the further away 3D topology and signal strength measurements. The CNN used is U-Net, which was first used for image segmentation and later for various other applications \cite{ronneberger2015u}. The U-Net architecture used is shown in Fig. \ref{fig:u-net} when $\frac{w}{d}=\frac{l}{d}=64$. The green and grey rectangles correspond to CNN layers with the numbers on top indicating the number of layers while the numbers on the side denote the size of each layer. Note that the rectangle with the number 2048 overlaying it is a single fully connected layer of size 2048. The output is flattened into a vector of size $\frac{wl}{d^2}=|\mathcal{Q}|$.   

\section{Dataset and algorithm training  }
\label{sec:dataset}
In this section we describe how we obtained the dataset used for training and testing of the algorithm. 

The main tool used for generating the dataset was the Wireless InSite ray-tracing software. The software takes in the 3D model of the environment, along with other parameters, such as transmission frequency, transmitter location and bandwidth, to trace the radio propagation paths and calculate the signal strength at the desired points. 

In order to create an expansive set of environments, we used a handcrafted script to generate Manhattan-grid-like urban environments. The script starts by dividing a rectangular area into city blocks with random dimensions. Then, open spaces and rectangular-base buildings are added within those blocks. 
The outputs from Wireless InSite were then processed in Python and used to train the algorithm.   

In total, we generated 45 urban environments of size 400m by 400m each. In each environment, we placed transmitters uniformly spaced at 80m apart, which equates to 25 transmitter positions per environment. However, if a transmitter location happens to be indoor following the said uniform spacing pattern, we removed that transmitter. For each transmitter position, in each environment, Wireless InSite was used to calculate the signal strength values over a grid of points spaced at $d$ at an altitude $h_{\text{UAV}}$. The calculations were ran for narrowband signals over a frequency of 800 MHz and the transmitted power was 20 dBm. 

The data generated in 30 out of 45 environments was used to train the algorithm while the data from the remaining 15 environments was used to evaluate it. The AoI dimensions selected were $w=l=240$m, hence to obtain $\mathbf{x}$, $\mathcal{M}$ and $\mathbf{y}$, only the data from a randomly selected $w\times l$ portion of an $400m \times 400m$ environment was sampled. We obtain $\mathbf{x}$ by selecting a random subset of entries of $\mathbf{y}$. We do not model any measurement error when generating $\mathbf{x}$. The spacing $d$ was 4m.   

We used the Adam optimization algorithm, which is an extension of the stochastic gradient descent algorithm to minimize the loss function in Eq. (1) \cite{kingma2014adam}. The default configuration parameters for Adam in Tensorflow were used. A constant learning rate of $10^{-5}$ with a batch size of 64 was used. The training takes about 6 hours to complete running on a machine with a GeForce GTX 1060 GPU.

\section{Results}
\label{sec:results}

\begin{figure}[t]
\centering
\begin{subfigure}{.24\textwidth}
  \centering
  \includegraphics[width=0.9\linewidth]{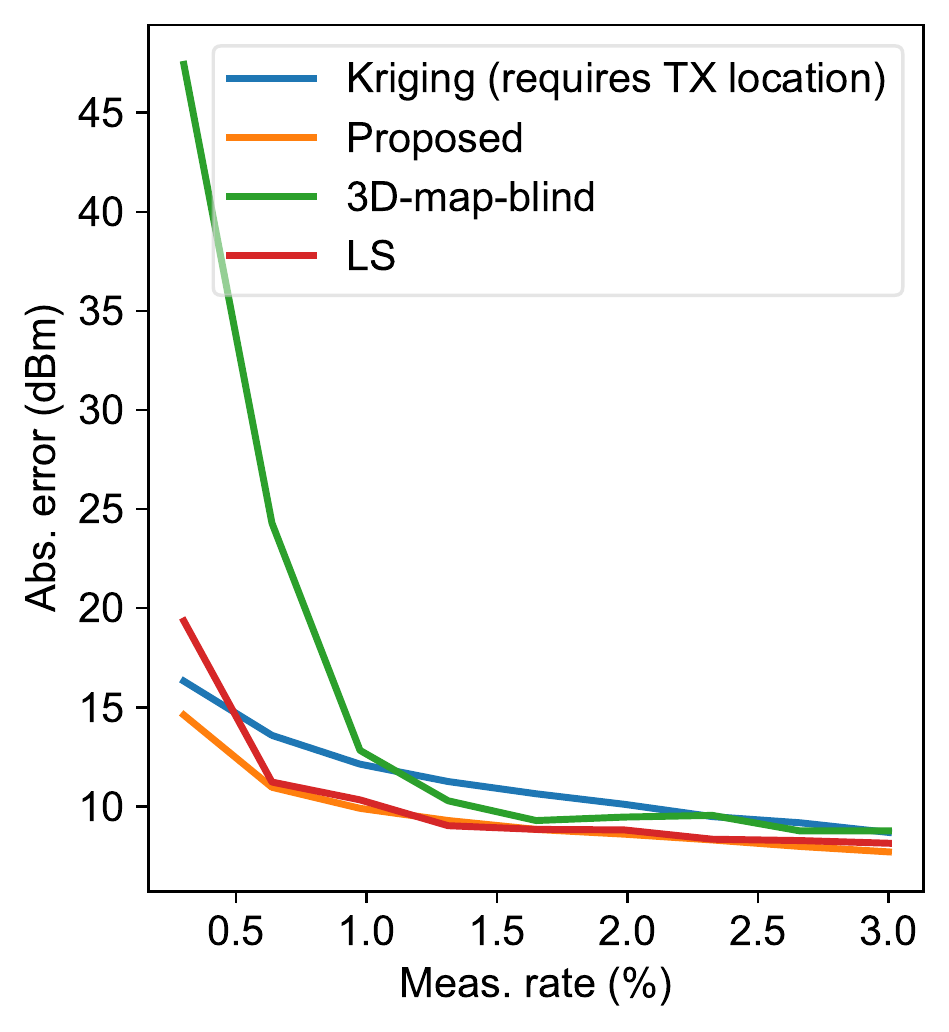}
  \caption{Random uniform samples}
  \label{fig:uniform-trend}
\end{subfigure}%
\begin{subfigure}{.24\textwidth}
  \centering
  \includegraphics[width=0.9\linewidth]{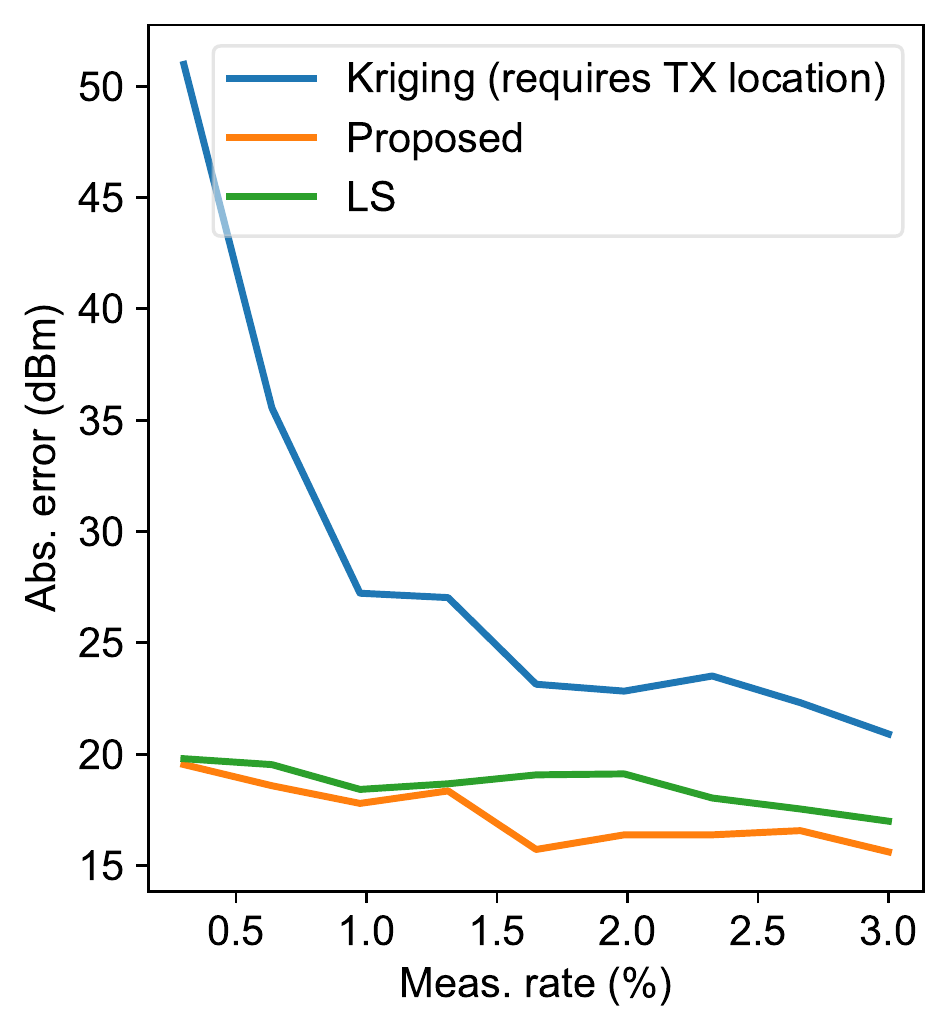}
  \caption{Random trajectory samples}
  \label{fig:random-trend}
\end{subfigure}
\caption{Average absolute error of our algorithm and the three benchmarks when the number of measurements is varied from 0.3 \% to 3 \% of $|\mathcal{Q}|$. }
\label{fig:trend}
\vspace{-10px}
\end{figure}

\begin{figure*}[t]
	\centering
	\includegraphics[width=0.95\textwidth]{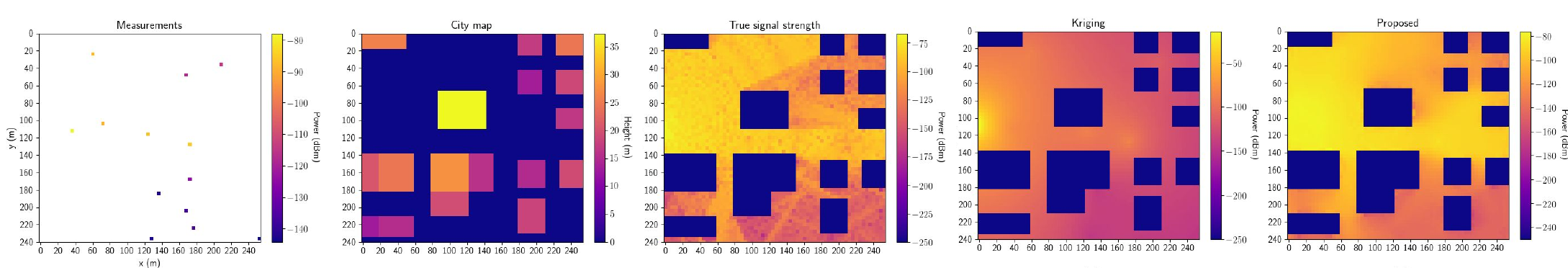}
	\caption{An example with random uniform placement of measurements. The average absolute error for the Kriging approach was 11.94 dBm while the error for the our approach was 6.95 dBm. }
	\label{fig:uniform}
	\vspace{-7px}
\end{figure*}

\begin{figure*}[t]
	\centering
	\includegraphics[width=0.95\textwidth]{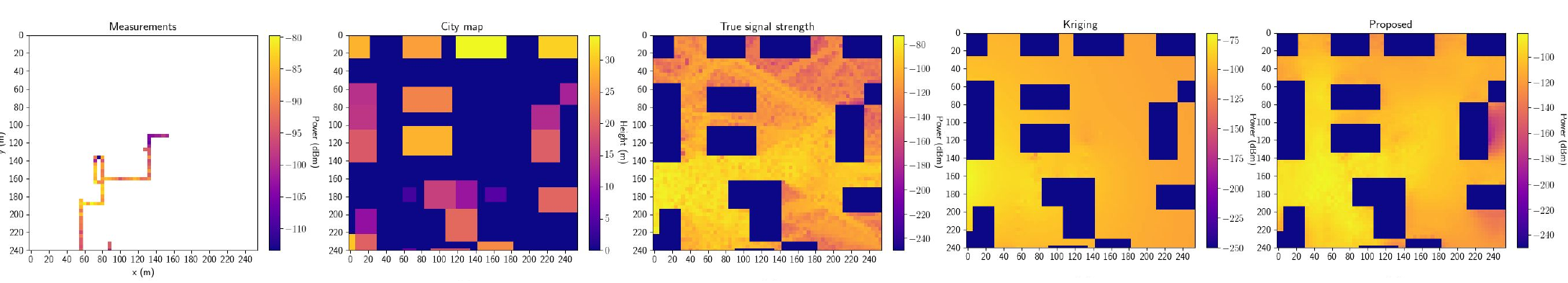}
	\caption{An example with measurements collected on a random trajectory. The average absolute error for the Kriging approach was 14.17 dBm while the error for the our approach was 11.11 dBm.}
	\label{fig:random-walk}
	\vspace{-10px}
\end{figure*}

In this section, we compare the performance of our algorithm on the testing data set against several benchmarks. 

We use three different benchmarks for comparison:
\begin{enumerate}
    \item 3D-map-blind approach: This approach is identical in every way to our proposed approach, except it only uses $\mathbf{x}$ as the input. By comparing the proposed algorithm against this approach we are verifying that our algorithm is successful at utilizing 3D maps and can arrive at a better prediction as a result. 
    \item Least-squares (LS) approach: This approach only predicts $\mathbb{E}[\tilde{\mathbf{y}}]$ and is identical to our approach in every other way. This approach only learns $\mathbf{\mu}_\theta(\mathbf{x}, \mathcal{M})$ and the loss function is a least squares regression against the true signal strength values: 
    $
    \mathcal{L}(\theta)=\frac{1}{N}\sum_{i=1}^{N}\Delta_{i}^{T}\text{diag}\left(\mathbf{z}_{i}\right)\Delta_{i}
    $. Of course, this approach is lacking since it does not predict the variance of its prediction, which is an important feature as we have explained earlier. 
    \item Kriging interpolation method: This is one of the most common approaches to spatial signal strength prediction. The details of the implementation of this approach can be found in \cite{chowdappa2018distributed}, for example, where a distributed and a centralized implementation are covered (we apply the centralized version). With this approach, the parameters of the path-loss and shadowing model are estimated from the collected measurements. Using these parameters, we determine the path-loss component of the channel gain. Then, with the path loss accounted for, we use Kriging spatial interpolation to estimate the shadowing. The Kriging approach cannot incorporate the use of 3D maps like our algorithm and it also requires knowing the location of the GU, therefore it is not an equal-grounds benchmark for out algorithm. Nevertheless, we include it due to its popularity and since we were not able to find any other equal-grounds comparison for our algorithm.
\end{enumerate}

We perform evaluation for two different methods of collection of signal strength measurements, $\mathbf{x}$:
\begin{enumerate}
    \item Uniform random placement of measurements: The samples are uniformly and randomly spread across the AoI. This would correspond to sensors randomly placed across the AoI, for instance. 
    \item Random-trajectory placement of measurements: This would correspond to a UAV moving across the AoI on a random trajectory and collecting the measurements. While a real UAV would not move on a random trajectory, we use a random trajectory to emulate measurements collection of UAV on some planned path. The random trajectory is essentially a random walk across the grid of the AoI where at each step the previous step is repeated at a probability $p$. This is done to add a momentum to the random movement and reduce the number of changes in direction of the movement.
\end{enumerate}
We describe the results for each method of sample collection separately.  

\subsection{Mean absolute error}
The first metric that we use for comparison is the mean absolute error: 
$
\frac{1}{|\mathbf{z}|}\left(\mathbf{y}-\mathbb{E}[\tilde{\mathbf{y}}]\right)^T \text{diag}\left(\mathbf{z}\right)
\left(\mathbf{y}-\mathbb{E}[\tilde{\mathbf{y}}]\right)
$.
The error is taken over only the outdoor locations.
The mean absolute error is measured over the 15 validation environments with different sampling of measurements.

\subsubsection{Uniform random placement of measurements}
We show the results for uniformly spaced measurements. In Fig. \ref{fig:uniform-trend} we vary the number of measurements expressed as the percentage of $|\mathcal{Q}|$ from 0.3 \% to 3 \%. Our proposed approach outperforms all three of the benchmarks. 

First, it has a slight edge over the LS approach at a higher number of measurements and a more significant advantage with a low number of measurements. This indicates that by learning to predict the distribution of the signal strength rather than just the expectation, the algorithm can perform well with a small number of measurements.    

Furthermore, the results in  Fig. \ref{fig:uniform-trend} show that our proposed algorithm particularly benefits from 3D map knowledge when there is only a small number of measurements available. 

Most importantly, our proposed algorithm outperforms the Kriging approach for all levels of collected measurements, even without knowing the transmitter location. Our algorithm does this by relying on 3D maps and a learning-based approach. 

We give a concrete example in Fig. \ref{fig:uniform} where we have a scenario with  0.5 \% measurements and compare our algorithm to the Kriging approach. For our algorithm, we only show the $\mathbf{\mu}_{\theta}(\mathbf{x}_{i},\mathcal{M}_{i})$ output. We purposely select a case with a small number of measurements since the performance of the two algorithms is similar with a high number of measurements. For scenarios with a small number of measurements, the Kriging algorithm is only capable of estimating the path-loss trend, while it gives little prediction of how the signal strength is perturbed by the NLOS components. Our algorithm gives an estimate which is much closer to the actual signal strength field. Using the building knowledge, it correctly predicts which areas will be occluded to the radio signals and which have better access to the signal.  

\subsubsection{Random-trajectory placement of measurements}

The benefit of our proposed algorithm is even more significant when samples are collected on a random trajectory across the AoI, as seen in Fig. \ref{fig:random-trend}. All four approaches perform worse compared to the uniform random placement of measurements, since the measurements are clustered closer together and it is more difficult to make inference across the entire AoI.  

In this case, we omit the results for the 3D-map-blind approach since its performance is far worse compared to other three approaches. It seems that the prediction of signal strength becomes more challenging when measurements are clustered together and without knowing the 3D map of the environment.  

Our proposed algorithm performs similarly to the LS approach, however the difference is more significant relative to the Kriging approach. The Kriging approach depends on measurements being correlated to where signal strength needs to be predicted, which cannot be guaranteed with measurements collected on a single path. 

An example case is given in Fig. \ref{fig:random-walk} with the number of measurements being 3\% of $|\mathcal{Q}|$. Our algorithm again gives a more nuanced prediction of the signal strength field even without knowing the transmitter location and correctly predicts which areas will be occluded from and which exposed to the signal. 

\subsection{Goodness of fit}
\begin{figure}[t]
\centering
\begin{subfigure}{.23\textwidth}
  \centering
  \includegraphics[width=0.9\linewidth]{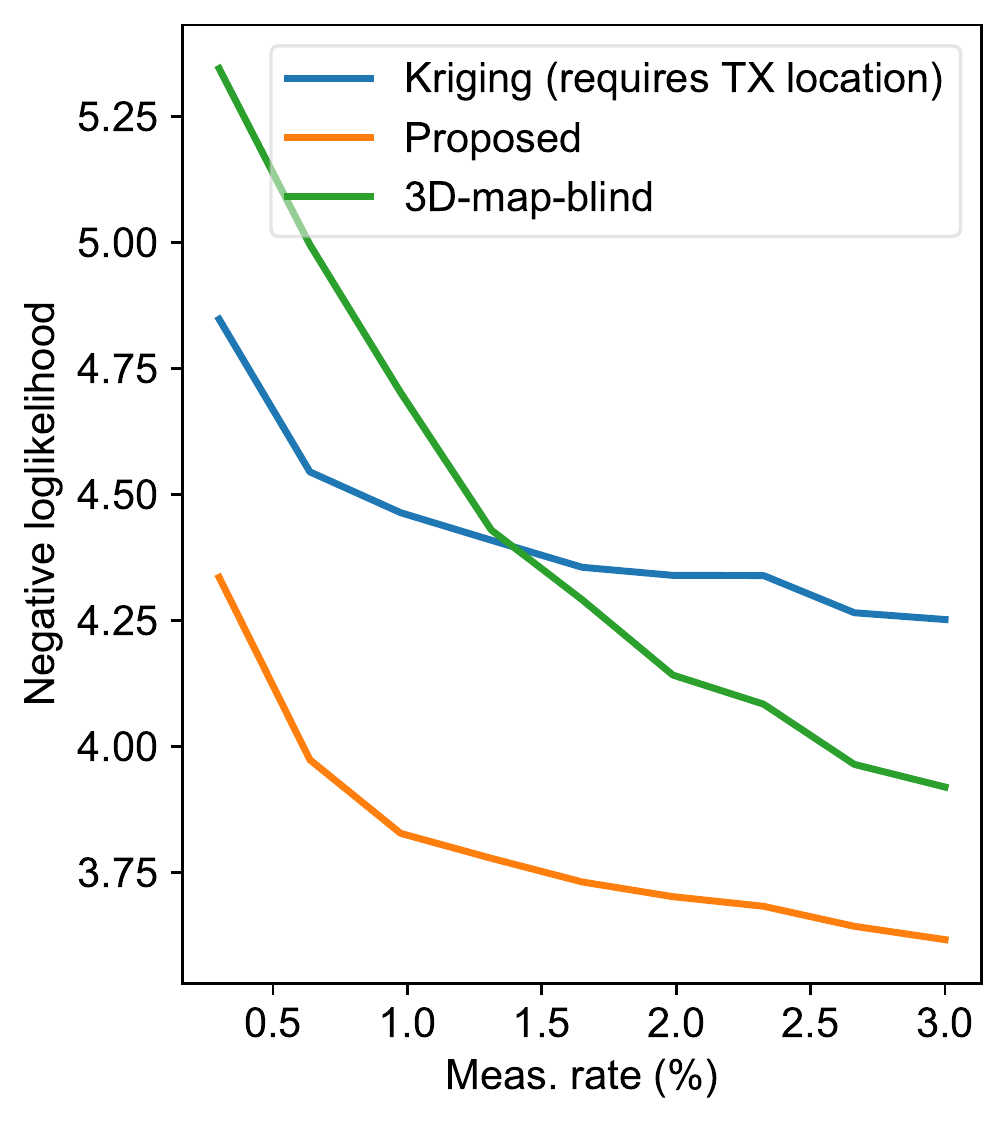}
  \caption{Random uniform samples}
  \label{fig:loglikelihood-uniform}
\end{subfigure}%
\begin{subfigure}{.23\textwidth}
  \centering
  \includegraphics[width=0.9\linewidth]{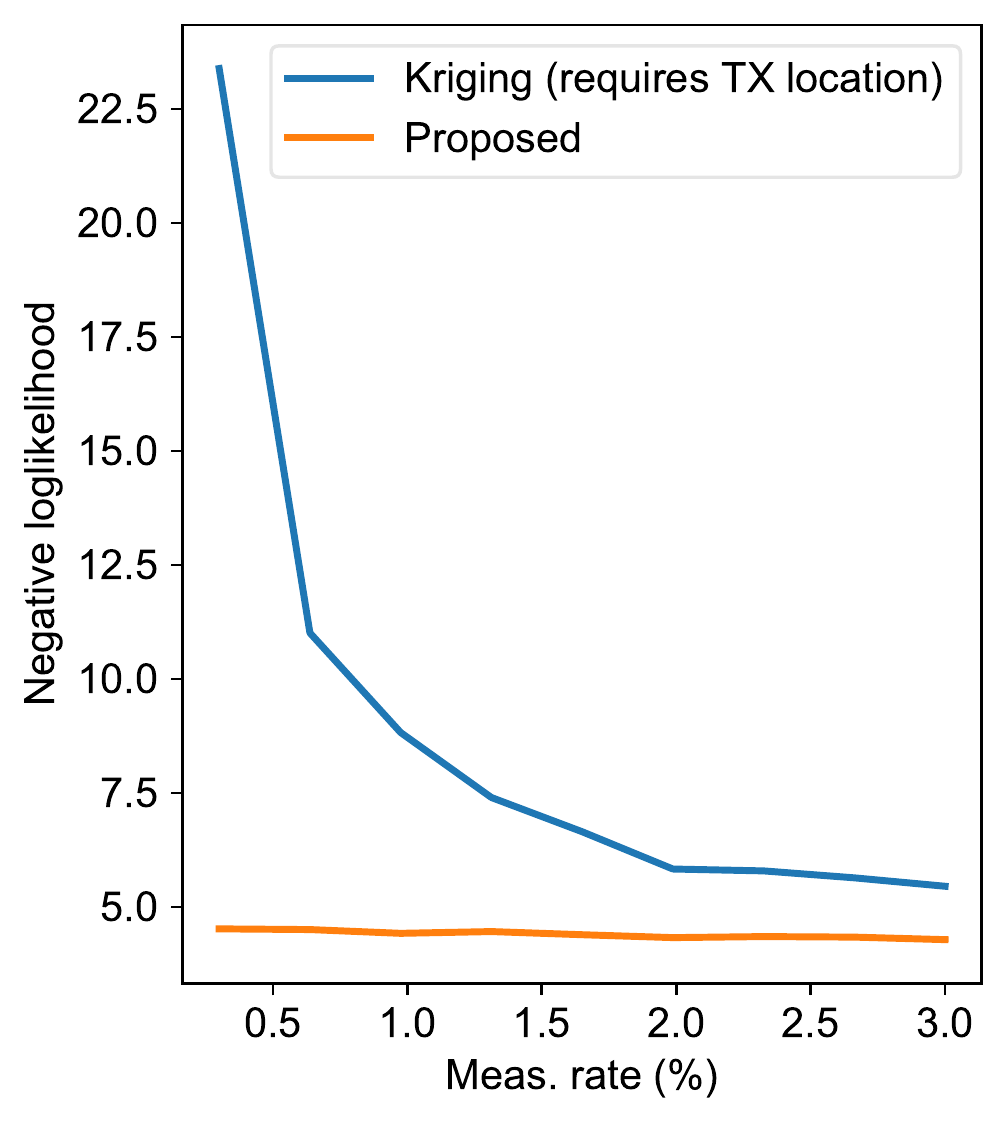}
  \caption{Random trajectory samples}
  \label{fig:loglikelihood-walk}
\end{subfigure}
\caption{Median log-likelihood of our algorithm and the two benchmarks when the number of measurements is varied from 0.3 \% to 3 \% of $|\mathcal{Q}|$. }
\label{fig:loglikelihood}
\vspace{-10px}
\end{figure}

To evaluate the goodness of fit of our approach, we measure the log-likelihood of the validation data-set with the predicted mean $\mathbb{E}[\tilde{\mathbf{y}}]$, and variance, $\text{Var}[\tilde{\mathbf{y}}]$. Since the distribution of the fitted variable is Gaussian, the negative log-likelihood is given by 
$
\frac{1}{2|\mathbf{z}|}\left(\mathbf{y}-\mathbb{E}[\tilde{\mathbf{y}}]\right)^T \text{diag}\left(\mathbf{z}\right) \text{diag} \left(\text{Var}[\tilde{\mathbf{y}}]\right)^{-1}
\left(\mathbf{y}-\mathbb{E}[\tilde{\mathbf{y}}]\right) +\\
\frac{1}{2|\mathbf{z}|}   \mathbf{1}^T\left( \text{diag} \left(\mathbf{z}\right)  \log  \left (\text{Var}[\tilde{\mathbf{y}}] \right)\right)    
$.
The log-likelihood is measured only over the outdoor locations. 
The Kriging approach also outputs mean and variance of the prediction, therefore we can treat the prediction as a Gaussian random variable and calculate the log-likelihood. 
The median negative log-likelihood across many Monte-Carlo iterations for the two sample collection methods is shown in Fig. \ref{fig:loglikelihood}.

Our algorithm achieves a lower median negative log-likelihood than both the Kriging and 3D-map-blind approach, which means that our approach achieves a better model fit to the test data. The fact that  negative log-likelihood of the proposed approach is lower than that of the 3D-map-blind approach suggests that our algorithm successfully takes advantage of the 3D map information. We omit the data for the 3D-map-blind approach for the random trajectory samples in Fig. \ref{fig:loglikelihood-walk}, since it performs much worse than the proposed or Kriging approach. Most importantly, the proposed algorithm achieves a better fit than the Kriging approach for all levels of collected measurements, with the improvement being drastic for measured samples collected on a random trajectory.

\section{Conclusions}
\label{sec:conclusions}

In this work, we have shown that deep learning can be used to predict radio signal propagation in an urban environment and we present a deep learning algorithm that can successfully accomplish this. Our algorithm successfully utilizes 3D maps to assist its prediction and provides a stochastic prediction of signal strength. It performs better than the benchmark solution for signal strength prediction, with the improvement being drastic when the prior signal strength measurements are collected on a path rather than being uniformly randomly placed. 
So far, our algorithm has only been evaluated in an environment simulated by a ray-tracing software and future work will involve a proof of concept with real outdoor measurements. 

\bibliography{references}
\bibliographystyle{ieeetr}

\end{document}